\documentclass[reprint,superscriptaddress,amsmath,amssymb,aps]{revtex4-1}

\usepackage{graphicx}
\usepackage{dcolumn}
\usepackage{bm}
\usepackage{gensymb}
\usepackage{color}
\usepackage[space]{grffile}
\usepackage{latexsym}
\usepackage{amsfonts,amsmath,amssymb}
\usepackage{url}
\usepackage[utf8]{inputenc}
\usepackage{fancyref}
\usepackage{hyperref}
\usepackage{float}
\usepackage{outlines}
\hypersetup{colorlinks=false,pdfborder={0 0 0},}
\usepackage{enumitem}
\usepackage{upgreek}

\usepackage{color,soul} 

\begin{document}
\title{A scalable multi-photon coincidence detector based on superconducting nanowires}
\author{Di Zhu}
\author{Qing-Yuan Zhao}
\author{Hyeongrak Choi}
\author{Tsung-Ju Lu}
\author{Andrew E. Dane}
\author{Dirk R. Englund}
\author{Karl K. Berggren}\email{berggren@mit.edu}
\affiliation{Research Laboratory of Electronics, Massachusetts Institute of Technology, Cambridge, Massachusetts, 02139, United States}
\date {November 27, 2017}

\begin{abstract}
Coincidence detection of single photons is crucial in numerous quantum technologies and usually requires multiple time-resolved single-photon detectors. However, the electronic readout becomes a major challenge when the measurement basis scales to large numbers of spatial modes. Here, we address this problem by introducing a two-terminal coincidence detector that enables scalable readout of an array of detector segments based on superconducting nanowire microstrip transmission line. Exploiting timing logic, we demonstrate a 16-element detector that resolves all 136 possible single-photon and two-photon coincidence events. We further explore the pulse shapes of the detector output and resolve up to four-photon coincidence events in a 4-element device, giving the detector photon-number-resolving capability. This new detector architecture and operating scheme will be particularly useful for multi-photon coincidence detection in large-scale photonic integrated circuits. 
\end{abstract}


\maketitle

\section{Introduction}
Single-photon detection plays a key role in quantum information processing, including modular quantum computing with trapped ions~\cite{Monroe2013} and solid-state quantum emitters ~\cite{Childress2014,Pant2017,Nemoto2014}, photonic quantum walks and Boson sampling~\cite{Peruzzo2010,Spring2013,Broome2013,Tillmann2013}, quantum simulations~\cite{Aspuru-Guzik2012}, and linear optical quantum computing~\cite{Knill2001}. Most of these applications rely on coincidence measurement of single or entangled photons over a large number of spatial modes and require an equal number of time-resolved single-photon detectors. Among various single-photon detectors~\cite{Hadfield2009}, the superconducting nanowire single-photon detector (SNSPD) has become increasingly attractive because of its outstanding detector metrics~\cite{Natarajan2012,Marsili2013,Najafi2015b,Marsili2012,Shibata2015} and feasibility of on-chip integration~\cite{Sprengers2011, Pernice2012, Reithmaier2015, Schuck2013, Najafi2015,Rath2015,Vetter2016}. Traditional SNPSD arrays used for space communication~\cite{Shaw2015},  photon number resolution~\cite{Dauler2006}, and few-channel coincidence counting~\cite{Najafi2015}  adopt parallel readout of individual detector elements. However, scaling these arrays for coincidence counting over large numbers of channels presents formidable challenges, especially for the electrical readout~\cite{Miki2011}.

A number of multiplexing schemes and device architectures have been developed to solve the readout problem. Row-column multiplexing is an efficient scheme but still requires $2N$ readout channels for $N^2$ pixels~\cite{Allman2015a}. Another promising scheme is the frequency-division multiplexing, where SNSPDs are embedded in resonators operating at different radio-frequency (RF) tones~\cite{Doerner2016, Doerner2017,Doerner2017a}. Though a common feed line can couple multiple resonators, each RF tone needs a demultiplexing circuit. Besides frequency-domain multiplexing, time-domain multiplexing has also been explored. Hofherr et al. demonstrated time-tagged multiplexing in a proof-of-concept two-element array~\cite{Hofherr2013}, in which the signals from the two elements were separated in time using a delay line. While this approach only required a single readout line,  the device dimension and array size were limited by the delay line design.  Time-tagged multiplexing was more recently employed by Zhao et al. to create a single-photon imager from a continuous nanowire delay line~\cite{Zhao2017}. This imager resolves photon position but was only used to detect one photon at a time. Another architecture connects nanowires in parallel and encodes the desired information in the amplitude of the electrical output, such as photon number~\cite{Divochiy2008} or position~\cite{Zhao2013}. However, these detectors require on-chip resistors for biasing, and the array size is limited by the leakage current to the parallel branches.

Here we report on a two-terminal detector based on superconducting nanowire microstrip transmission lines that works as a scalable array. Unlike previous work~\cite{Hofherr2013,Zhao2017}, this detector resolves the location of more than one photon and works naturally as a coincidence counter. With simple timing logic, we demonstrated the resolution of all 136 possible single- and two-photon events in a 16-element detector. With pulse shape processing, we resolved up to four-fold coincidence events and showed photon-number-resolving capability in a 4-element device. The microstrip transmission line used in the detector had a group velocity as low as 1.6\%$c$ (where $c$ is the speed of light in vacuum) and may allow denser packing compared to co-planar structures~\cite{Zhao2017}. The detector was designed for integration on optical waveguide arrays and fabricated on a waveguide-compatible substrate material. We expect it to find immediate applications in large-scale on-chip coincidence detection for quantum information processing.
	
\section{Results}

\begin{figure*}
	\includegraphics{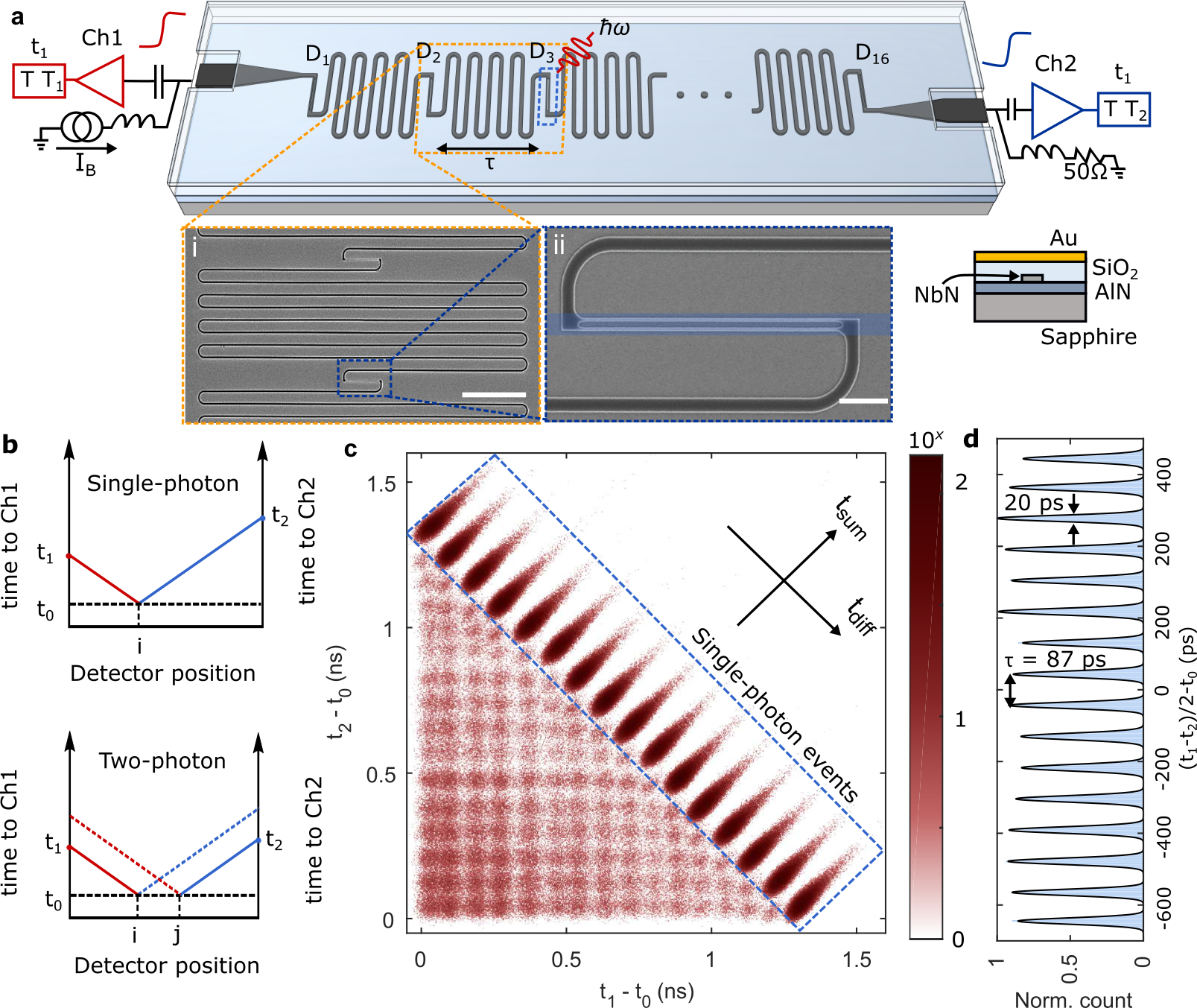}
	\caption{\textbf{Device architecture and operating mechanism.} \textbf{a}, Device layout for a 16-element two-photon detector. The two-photon detector is a two-terminal array that connects a chain of single-photon detectors using slow-wave nanowire delay lines. The nanowire was designed to be a microstrip waveguide with a dielectric spacer and top ground plane. (i): Scanning electron micrograph (SEM) of the delay line, which consisted of a 300-nm-wide meandered nanowire. Scale bar: 10\,$\upmu$m. (ii) SEM of a detector segment, which consisted of two 80-nm-wide parallel nanowires. Scale bar: 1\,$\upmu$m. The blue band marks the site for future waveguide integration. \textbf{b}, Illustration of the timing logic in the detector. $t_0$ is the photon arrival time; $t_1$ and $t_2$ are the times when the electrical signal arrives at Ch1 and Ch2, respectively. TT: time tagger.  \textbf{c}, Measured timing distribution that resolved all 136 distinguishable groups. The color bar is in log scale, and the time bin is $\sim$3\,ps. \textbf{d}, A 1-D histogram of the difference time ($(t_1-t_2)/2$) when the detector was operating in the single-photon regime. The efficiency of the segments was uniform with a standard deviation of 6\% of its mean value. The average FWHM differential jitter was 20\,ps and timing delay between adjacent segments was 87\,ps.}
    	\label{fig:fig_1}
\end{figure*}
    
\paragraph{Device architecture.}
Fig.~\ref{fig:fig_1} illustrates the basic architecture and operating principle of the two-photon detector. In our design, individual detecting elements were connected by nanowire delay lines, resulting in a one-dimensional detector array. Figure~\ref{fig:fig_1}a shows the physical implementation of a 16-element ($D_1$ to $D_{16}$) array. The fabrication is described in the Methods Section. The detectors and delay lines were patterned from a $\sim$5-nm-thick niobium nitride (NbN) film on aluminum nitride (AlN)-on-sapphire substrate~\cite{Zhu2016}. Figure~\ref{fig:fig_1}a panel (i) shows a scanning electron micrograph (SEM) of the delay line. Each meandered delay line had a width of 300 nm, a period of 1.8\,$\upmu$m, and a total length of 429\,$\upmu$m. Figure~\ref{fig:fig_1}a panel (ii) shows an SEM of a detector segment. Each detector segment consisted of a pair of 80-nm-wide, 5-$\upmu$m-long parallel nanowires. This detector segment design is known as a 2-element superconducting nanowire avalanche photodetector (2-SNAP)~\cite{Ejrnaes2007}. Compared to a standard hairpin nanowire~\cite{Pernice2012}, the 2-SNAP enhanced signal-to-noise ratio and provided relatively good impedance matching to the 300-nm-wide delay lines. To make the nanowires into transmission lines, we capped the device area with a 450-nm-thick oxide spacer and 60-nm-thick gold ground plane on the top (see Fig.~\ref{fig:device_micrograph} in the Supplementary Information for micrographs of the overall device). Designing the isolated nanowires as transmission lines was essential for the delay-line-based detector: the transmission line guided the RF signal along the nanowire with a slow propagation speed and minimized RF coupling in the meander. The current device design was intended for future integration with a 16-channel AlN photonic waveguide array (see Supplementary Information). The blue shaded band in Fig.~\ref{fig:fig_1}a panel (ii) marks the potential position for an optical waveguide.

The detector was biased using a constant DC current and read out on both terminals (Ch1 and Ch2) using room-temperature low-noise amplifiers~\cite{Calandri2016, Zhao2017}. When the 2-SNAP was biased close to its critical current, the delay line was only biased at $\sim$50\% and therefore would not respond to incident photons. 

Figure~\ref{fig:fig_1}b illustrates the timing logic in the detector. In the single-photon regime (see the upper panel in Fig.~\ref{fig:fig_1}b), only one segment fires at a time, following the timing logic as presented in Ref.~\cite{Zhao2017, Hofherr2013}. For instance, if a photon arrives on the $i$-th detector ($D_i$) at time $t_0$ and excites a pair of counter-propagating pulses, the left-propagating pulse will reach Ch1 at time $t_1=t_0+(i-1)\tau$, where $\tau$ is the delay between two adjacent segments; and the right-propagating pulse will reach Ch2 at time $t_2=t_0+(N-i)\tau$, where $N$ is the number of segments in the array. In this case, the arrival time of the photon can be derived from the sum of the two pulse times,  $(t_1+t_2)/2=t_0+(N-1)\tau/2$, while the arrival location of the photon is determined from their difference, $(t_1-t_2)/2\tau=i-(N+1)/2$. 

The timing logic is different for the two-photon case (see the lower panel in Fig.~\ref{fig:fig_1}b). When two segments fire at the same time, each of them launches a pulse pair, but each readout channel will only identify the pulse edge from their nearest segment because the pulse width (ns) is significantly larger than the delay time (ps). So if $D_i$ and $D_j$ both fire ($i<j$), Ch1 will tag $t_1=t_0+(i-1)\tau$, while Ch2 will tag $t_2=t_0+(N-j)\tau$. If $t_0$ is known, one can trace back both $i$ and $j$. This method requires the knowledge of $t_0$, which is available in many practical applications. For pulsed single-photon or photon-pair sources, the excitation laser gives $t_0$; in communication or computing, the reference clock gives $t_0$ as long as the timing window and timing jitter is smaller than $\tau$. 

\paragraph{Measurement result in a 16-element detector.} Figure~\ref{fig:fig_1}c shows the measured timing distribution in a 16-element detector. 136 groups of detection events can be distinguished. The diagonal groups correspond to the 16 single-photon detection cases, and the off-diagonal groups correspond to the 120 ($C^{16}_2$) two-photon detection cases. Like all array-type photon-number-resolving detectors, the cases where two photons hit the same detector (with probability  $O(1/N)$) cannot be resolved. The observed higher counting rate at $t_1-t_0$ and $t_2-t_0$ near zero (lower-left corner in the histogram) was due to the increasing probability of hidden multi-photon detection (more than two segments fire simultaneously).   The histogram was constructed from 1 million detection events, discretized here in bins of $\sim$3\, ps. The detector was measured at 3.0\,K and flood-illuminated from the back of the chip using a 1550\,nm sub-ps pulsed laser. It was biased at 14.5\,$\upmu$A with a switching current of 15.3\,$\upmu$A.

It is useful to introduce two characteristic timing variables: the sum time, $t_\mathrm{sum} = (t_1+t_2)/2-t_0$, and the difference time, $t_\mathrm{diff}=(t_1-t_2)/2$. ($t_\mathrm{sum}$, $t_\mathrm{diff}$) forms a basis that is rotated relative to the ($t_1$-$t_0$, $t_2$-$t_0$) basis by 45\degree . As illustrated in the space-time diagram shown in Fig.~\ref{fig:fig_1}c, in the single-photon regime, $t_\mathrm{diff}$  reveals the segment position, while $t_\mathrm{sum}$ is a constant regardless of the position. 

To characterize the delay line and the uniformity of each detecting segment, we operated the detector in the single-photon regime and constructed a 1-D histogram for the difference time (see Fig.~\ref{fig:jitter_16_element} for the sum time). As shown in Fig.~\ref{fig:fig_1}d, the difference time histogram consisted of 16 Gaussians. The full-width at half-maximum (FWHM) was 20.3$\pm$0.6\,ps (average value with 1 $\sigma$ uncertainty), and the standard deviation of the peak amplitude was 6\% of its mean.

The 429-$\upmu$m-long delay line between each detector created an 86.8$\pm$0.3\,ps delay, corresponding to a signal propagation speed of 1.6\%$c$. The anomalously slow group velocity was due to the high kinetic inductance of the superconducting nanowire and large capacitance offered by the top ground plane placed 450\,nm above the nanowire~\cite{Pippard1947}. Instead of a full field solution~\cite{Mason1969, Swihart1961, Mason1969, Chang1979}, the characteristic impedance and phase velocity of our nanowire transmission line was estimated using a distributed circuit model, where $Z_0 = \sqrt{L_\mathrm{s}/C_\mathrm{s}}$ and $v_\mathrm{p}=1/\sqrt{L_\mathrm{s}C_\mathrm{s}}$. Here, $L_\mathrm{s}=L'_\mathrm{K}+L'_\mathrm{F}\approx L'_\mathrm{K}$, where $L'_\mathrm{K}$ and $L'_\mathrm{F}$ are the specific kinetic and Faraday inductances, respectively; and $C$ is the specific capacitance. From numerical simulation, we estimated $L_\mathrm{s}\approx$0.3\,mH/m (0.3\,nH/$\upmu$m) and $C_\mathrm{s}\approx$128\,pF/m (0.128\,fF/$\upmu$m).

\begin{figure*}
	\includegraphics{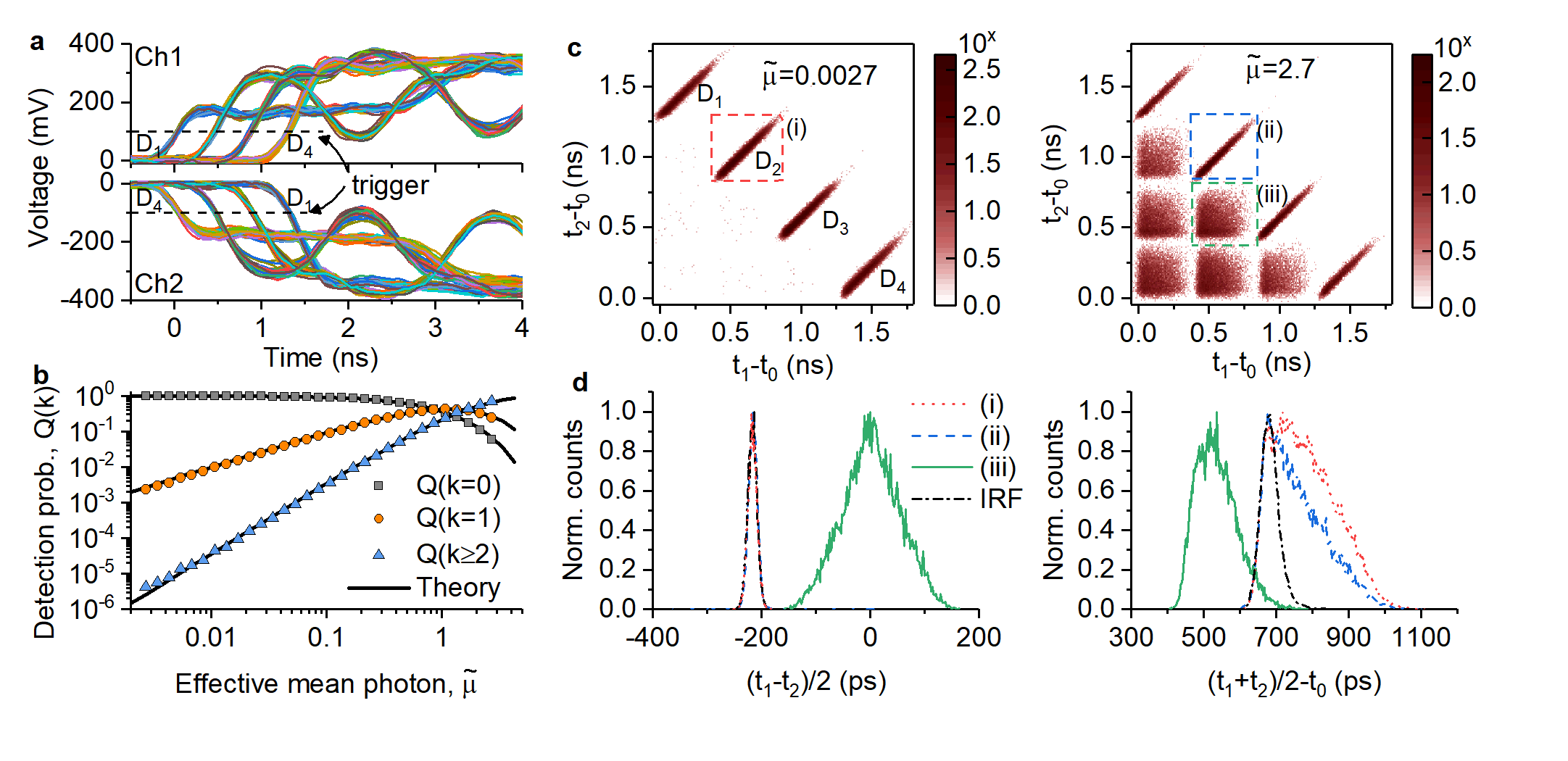}
	\caption{\textbf{Timing logic based two-photon detection in a 4-element detector chain.} \textbf{a}, Electrical pulses from a 4-element device when illuminated with a sub-ps pulsed laser in the single-photon regime. The dotted lines mark the trigger level for time tagging. \textbf{b}, Photon counting statistics under different input powers. $\tilde{\mu}=\eta\mu$ is the effective mean photon per pulse which included detector inefficiency. $Q(k)$ is the probability that $k$ segments fire. The symbols are measurement results, and for comparison, the lines are calculated from a theoretical model based on the detector conditional probability and the Poisson distribution of the coherent state input.  \textbf{c}, Measured timing distribution for $\tilde{\mu}=0.0027$ (left) and $\tilde{\mu}=2.7$ (right). The time bin in the plot is 10 ps and color bar is in log scale.  \textbf{d}, Histogram of the difference (left) and sum (right) time for 4 representative groups of detection events. (i), (ii): $D_2$ fires under weak and strong illumination, respectively; (iii): $D_2$ and $D_3$ fire simultaneously. (i-iii) are labeled in \textbf{b} and were measured using a laser with pulse width $\sim$200\,ps. IRF: instrument response function of D$_2$ probed using a sub-ps pulsed laser in the single-photon regime, showing a FWHM differential jitter of 20\,ps and sum jitter of 56\,ps.}
	\label{fig:fig_2}
\end{figure*}

\paragraph{Analysis in a 4-element detector.} For clarity, we performed detailed timing and photon counting analysis on a widely separated 4-element ($D_1$-$D_4$)detector array. The 4-element detector had the same design as the 16-element device but with a five times longer delay between adjacent detector segments.

Figure~\ref{fig:fig_2}a shows 200 pairs of electrical pulses from the detector when illuminated using a sub-ps pulsed laser in the single-photon regime. The pulses were aligned according to the timing reference from the laser. The dashed line marks the trigger level for time tagging, where the four groups of pulses were separated by $\sim$426\,ps.

Depending on the position of the firing segment, the output pulse shapes were different. This position dependence was due to signal reflections in the nanowire. Besides the major impedance mismatch between the nanowire (1.5\,k$\Omega$) and the readout (50\,$\Omega$), the resistive hotspot (a dynamic resistance on the k$\Omega$ scale) also contributed to reflections. The reflections caused distinct pulse shapes for each detection event. For instance,  the pulses from $D_1$ on Ch1 had two rising edges separated by $\sim$3\,ns, which matched the round trip time  in the nanowire. Due to symmetry, pulses from $D_4$ on Ch2 also had the same signature. Impedance matching tapers could in principle be used to minimize reflections, enhance signal levels, and provide faster rising edges to reduce timing jitter~\cite{Zhao2017}. In our case, instead of performing a perfect impedance matching with a centimeter long taper, we used a short taper with high cut-off frequency. Though the imperfect impedance matching resulted in large reflections, it was possible to trigger at a lower threshold to capture only the initial part of the rising edge. Also, as will be shown later, the distinctive pulse shapes caused by reflection actually enabled us to resolve more than two photons.

\paragraph{Photon counting statistics.} We demonstrated the detector's ability to resolve single- and two-photon events by performing a photon-statistics measurement of a coherent source. The measured photon statistics $Q(k)$ are related to the source distribution $S(m)$ by $Q(k)=\sum_{m=0}^{\infty} P(k|m)S(m)$, where $P(k|m)$ is the conditional probability that $k$ detector segments click given $m$ photons in the source. The laser diode serving as the  input in our experiment follows the Poisson distribution,  $S(m)=\sum_0^\infty \frac{\mu^m}{m!}e^{-\mu}$, where $\mu$ is the mean photon number. Figure~\ref{fig:fig_2}b shows the measured $Q(k)$ when the  effective mean photon per pulse of the input laser $\tilde{\mu}$ was attenuated from 2.7 to 0.0027 using a calibrated variable attenuator. The measurement result (symbols) matched our theoretical model (lines, see Method section for the derivation). Here, $\tilde{\mu}=\eta\mu$ included detector and coupling inefficiencies.  The value of $\tilde{\mu}$ was estimated by fitting the measured zero-photon probability to $e^{-\tilde{\mu}}$ based on the known attenuation value.  For each mean photon number, we accumulated 100,000 detection events (not including non-click events)  and extracted the one- and two-photon detection probabilities using the timing logic. The zero-photon  probability was measured separately by counting the number of non-click events over 50,000 photon pulses. Doing so ensured enough samples for low probability events and minimized measurement shot noise, while avoiding the unnecessarily large number of measurements for high probability events.

\paragraph{Timing resolutions.} 

Figure~\ref{fig:fig_2}c shows the timing distribution for $\tilde{\mu}=0.0027$ (left panel) and $\tilde{\mu}=2.7$ (right panel). When $\tilde{\mu}=0.0027$, the detector was operating in the single-photon regime, and only the 4 diagonal groups were present. When $\tilde{\mu}=2.7$, the 6 off-diagonal groups became prominent. Here, each 2-D histogram was constructed from $\sim$100,000 detection events. In these measurements, the probing laser had a FWHM pulse width of $\sim$200\,ps. Therefore, the spread of each detection group in the 2-D histogram was significantly wider than that shown in Fig.~\ref{fig:fig_1}c. 

The spread of the timing distribution was affected by both the device timing jitter as well as the laser pulse width. As shown in Fig.~\ref{fig:fig_2}c, the single-photon events, compared to the two-photon events, had a slimmer distribution in the $t_\mathrm{diff}$ axis. The timing uncertainty  for each time tag consists of 3 parts: $\sigma^2_{t_\mathrm{1,2}-t_0} = \sigma^2 _\mathrm{ph} + \sigma^2_\mathrm{det} + \sigma^2_\mathrm{e}$, where $\sigma_\mathrm{ph}$ is the photon arrival jitter, i.e. the photon could hit the detecting segment at anytime in the optical pulse duration; $\sigma_\mathrm{det}$ is the detector intrinsic jitter, i.e. the absorbed photon could trigger a voltage pulse with a variable time; and $\sigma_\mathrm{e}$ is the electrical jitter, i.e. the electrical noise would fluctuate the trigger point on pulse rising edge~\cite{Sidorova2017, Zhao2011}. 

We extracted the timing distributions for 4 representative groups of detection events and compared them in Fig.~\ref{fig:fig_2}d. The 4 groups are (i) weakly illuminated single-photon detection on $D_2$, (ii) strongly illuminated single-photon detection on $D_2$, (iii) two-photon detection where $D_2$ and $D_3$ both fire, and (iv) single-photon detection on $D_2$ probed using a sub-ps pulsed laser instead of a $\sim$200 ps modulated laser diode. The last group is labeled as IRF (instrument response function) in the figure, because the laser pulse width had a negligible contribution to the measured timing jitters. For all of the single-photon detection events, the differential timing jitter (left panel) only contained the electrical jitter since both the photon arrival jitter ($\sigma_\mathrm{ph}$) and detector intrinsic jitter ($\sigma_{\mathrm{det}}$) were canceled  (see curves i, ii, and IRF). The measured FWHM differential jitter here was 20\,ps. For the two-photon detection cases, however, two segments could absorb photons at different times due to the finite optical pulse width, so the differential jitter also contained the photon arrival jitter ($\sigma_\mathrm{ph}$, see curve (iii)).  For the sum jitter (right panel), the IRF shows an intrinsic FWHM sum jitter of 56\,ps, which was primarily electrical jitter and detector intrinsic jitter. This value is consistent with our previous result in a NbN SNSPD on AlN substrate~\cite{Zhu2016}. It is noticeable that under strong illumination (ii and iii), the sum jitter became narrower compared to that in the weak illumination case (i). This reduction was likely due to the higher probability of photon absorption in the early part of the optical pulse when the pulse energy increased.

\begin{figure}
	\centering
	\includegraphics{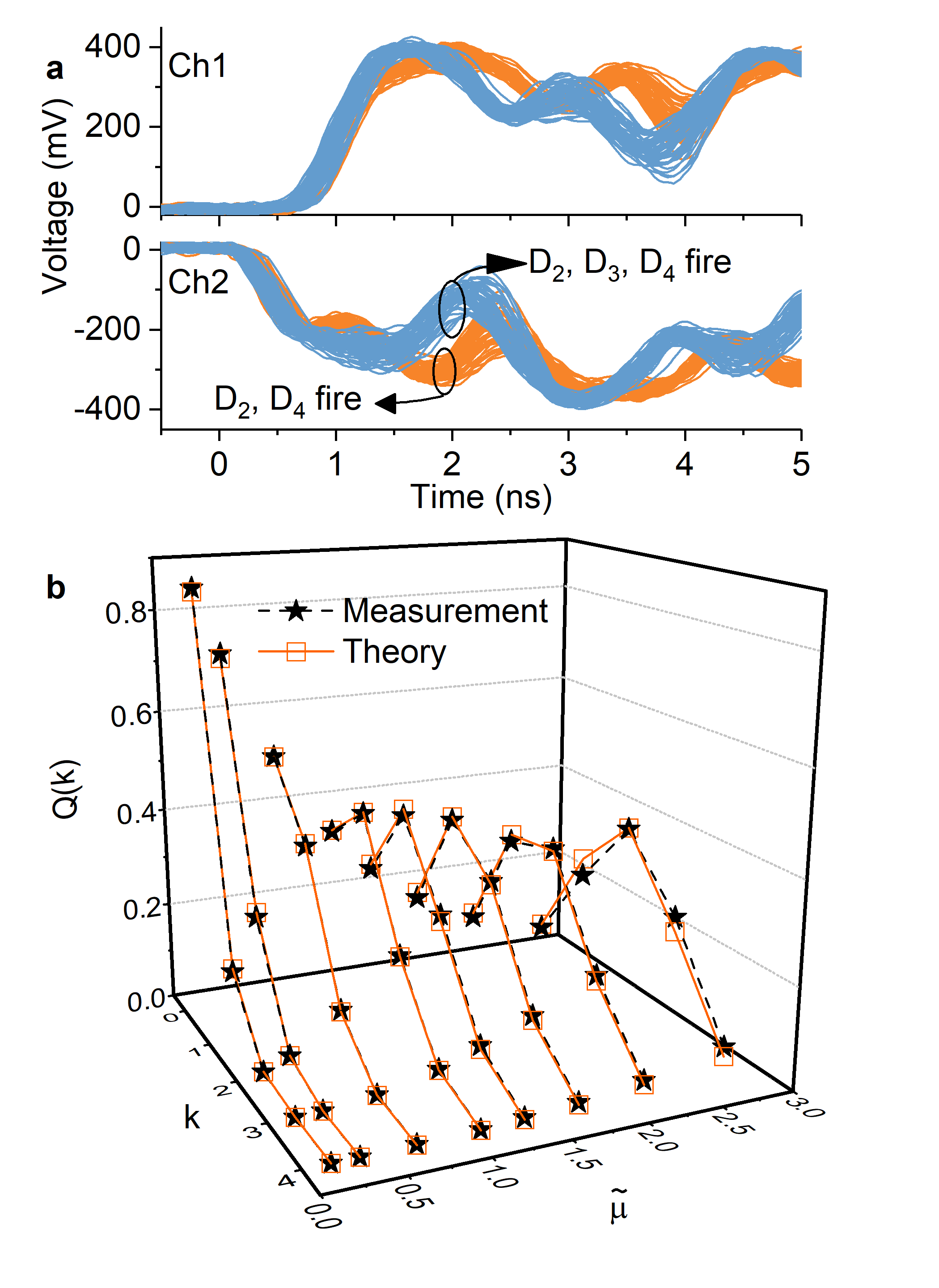}
	\caption{\textbf{Resolving more than two photons based on pulse shape processing.} \textbf{a}, Traces of 100 pairs of detector pulses, corresponding to detection events where $D_2$ and $D_4$ fired, or $D_2$, $D_3$, and $D_4$ fired. These events are indistinguishable based on timing logic since they have identical timing tags, but can be resolved from their pulse shapes. \textbf{b}, Measured photon statistics $Q(k)$ for $k$ up to 4 under coherent source illumination with mean photon per pulse $\tilde{\mu}$ from 0.27 to 2.7. 50,000 pairs of detection pulses were processed for each mean photon number, and the error of each data point was limited by the measurement shot noise. }
	\label{fig:fig_3}
\end{figure}

\paragraph{Beyond two-photon detection.} 
Two time tags can only resolve up to two-photon events. When 3 segments fire simultaneously, each readout channel will only register the rising edge from their nearest segment, and the signal from the middle segment will not be time tagged. For instance, when $D_2$, $D_3$, and $D_4$ fire together, it will produce the same time tags as $D_2$ and $D_4$ firing simultaneously. To resolve the difference, we need to process the detailed electrical pulse shapes. 

Figure~\ref{fig:fig_3}a shows traces of 100 pairs of detector pulses from the cases where $D_2$ and $D_4$ fired (orange traces) or $D_2$,  $D_3$, and $D_4$ fired (blue traces). These pulses have distinct signatures that allow them to be distinguished (e.g. the opening eye marked in Fig.~\ref{fig:fig_3}a). In principle, due to the reflections in the nanowire,  each detection event will have its own fingerprint in the output pulse shape. By learning and discriminating  these pulse shapes, one can resolve all the events without ambiguity.

In Fig.~\ref{fig:fig_3}b we demonstrated the capability of resolving up to four-photon events using the 4-element detector. The input optical field was attenuated from an effective mean photon per pulse of 2.7 to 0.27. For each measured attenuation level, we acquired  50,000 pairs of detector pulses. By analyzing the pulse shapes, we discriminated all $2^4-1$ combinations of detection events and sorted them into one-, two-, three-, and four-photon events. A complete list of all of the output pulses and their fingerprints can be found in the Supplementary Information. The zero-photon probability was measured in the same way as in Fig.~\ref{fig:fig_2}b. 

\section{Discussion}

The detector in our experiment was broadband responsive and had saturated internal quantum efficiency at 780\,nm wavelength (see Fig.~\ref{fig:DE_DCR} for efficiency characterization at different photon energies). Based on previous results~\cite{Zhu2016}, a 60-nm-wide 2-SNAP using the same material and substrate can  saturate at 1550\,nm. The optical absorption can in principle approach unity when the detector is integrated onto an photonic waveguide~\cite{Hu2009a,Pernice2012}. More details on waveguide integration and optical absorption can be found in the Supplementary Information.

The microstrip architecture used here offers significant advantages. When used as a free-space or fiber-coupled detector,  the ground plane and dielectric spacer can form an optical cavity to enhance absorption~\cite{Rosfjord2006}. Compared to co-planar waveguides, the microstrip can be meandered with a higher fill-ratio without having light-absorbing ground plane around the nanowire, which is suitable for high-efficiency single-photon imagers~\cite{Zhao2017}. 

The number of segments in the detector can be increased without additional biasing/readout resources. However, the maximum counting rate will decrease due to the kinetic inductance limit ~\cite{Kerman2006a}. Our current 16-element detector had a maximum counting rate of 4.8 \,MHz (see Supplementary Information for more discussion). With increasing segments, the timing logic remains simple, but the pulse shape analysis may become challenging. We are currently building physical and mathematical models to simulate and understand the detailed pulse shapes in the detector. 

In conclusion, we have developed a scalable coincidence detector based on superconducting nanowires. We engineered the nanowire to a microstrip transmission line with a group velocity as low as 1.6\%$c$. By varying the width at different sections, the nanowire serves either as a photon-sensitive detector segment or a compact delay line.  The timing-logic operation is ideal for two-photon coincidence counting over large numbers of spatial modes, while the pulse-shape processing enables higher-order coincidence measurements.  The device architecture is suitable for integration on optical waveguides and cavities. With increasing number of detecting segments, we expect it to provide a practical solution for implementing large-scale photonic quantum information processing systems. 

\section{Methods}

\paragraph{Detector fabrication.}
The NbN film was deposited on an AlN-on-sapphire substrate (Kyma Technologies, Inc.) using DC magnetron sputtering at 840 \degree C. The NbN deposition and nanowire patterning follows that described in Ref.~\cite{Najafi2015b}. The AlN was c-axis oriented with a thickness of 200$\pm$5\% nm. The NbN film had a thickness of $\sim$5 nm, critical temperature of 10.7 K, transition width of 1.63 K, sheet resistance of 510\,$\Omega$/sq, and residual resistance ratio of 0.85. The gold contact pads were patterned using photolithography followed by metal evaporation (5 nm Ti/50 nm Au/5 nm Ti) and lift-off. The superconducting nanowires were patterned using electron-beam lithography (Elionix F125) with a negative-tone resist, hydrogen silsesquioxane (HSQ), and etched using CF$_4$ reactive-ion etching. The dielectric spacer was fabricated by spin-coating the sample with a flowable Oxide (Dow Corning FOX-16) and curing the intended area with electron beam exposure. The thickness of the oxide spacer was measured to be 450\,nm using a surface profiler (Dektak). The top grounding plane was fabricated with an aligned photolithography followed by metal evaporation and lift-off (5 nm Ti and 60 nm Au). See Fig.~\ref{fig:fab_flow} for the detailed process flow.

\paragraph{Detector measurement.}
All measurements were performed in a pulse-tube-based cryostat at 3.0\,K. Each detector was wire bonded to a printed circuit board and connected to room temperature readout circuits through a pair of SMP cables (Ch1 and Ch2). The DC bias current was injected from Ch1 using a bias tee. The RF signal from each channel was amplified using a low-noise amplifier (MITeq AM-1634-1000, gain: 50 dB, bandwidth: 50\,kHz-1G\,Hz ) and acquired using a 6\,GHz real-time oscilloscope  (Lecroy 760Zi) or counted using a 22\, MHz universal counter (Agilent 53132A).  The detector chip was back illuminated through a single-mode optical fiber (SMF-28e). The fiber was mounted on a piezo-stage (Attocube) for alignment and focusing. When measuring the 16-element detector and probing the intrinsic timing response of the 4-element detector, a sub-picosecond fiber-coupled mode-locked laser (Calmar FPL-02CCF) was used. It has a center wavelength of 1550 nm and repetition rate of 20 MHz. During the experiment, the repetition rate was reduced to 500 kHz using an electro-optic modulator. When measuring the multi-photon response of the 4-element detector, a 1550 nm modulated diode laser was used (PicoQuant LDH-P-C-1550 laser head with PDL 800-B driver). The pulse was asymmetric, non-Gaussian, with a width of $>$200\,ps (see Fig.~\ref{fig:laser_pulse_width} in the Supplementary Information for the pulse shape estimation). The repetition rate was set to 1 MHz. In both cases, the laser output was split into two paths, one to a fast photodetector (Thorlabs DET08CFC) as the timing reference, and other to the detector with a calibrated variable attenuator (JDSU HA9) and a polarization controller.  More details on the characterization of standard detector metrics can be found in the Supplementary Information.

\paragraph{Derivation for photon counting statistics.}
In the generic case, when an optical mode illuminates on an $N$-element detector, each photon has probability $c_i$ of reaching detector $D_i$, which has an detection efficiency $\eta_i$. $c_i$ depends on the spatial mode of the input field, while $\eta_i$ is intrinsic to the detector. To simplify the modeling, we assumed a uniform detection efficiency for all elements (i.e. $\eta_i=\eta_j=\eta$). This assumption is reasonable based on our experimental characterization. We measured the detection efficiency distribution by driving the probing fiber far away from the device and uniformly illuminating the detector. For both the 4-element and 16-element detector chain, the standard deviation in $\eta_i$ was $<6\%$ of its mean (see Fig.~\ref{fig:fig_1}e and Fig.~\ref{fig:counting_ratio}). Under this assumption, we treated each segment as a perfect detector with unity efficiency and incorporated the detector inefficiency to the input field, which makes the input mean photon number to be $\tilde{\mu}={\eta\mu}$. Here, we also included coupling inefficiencies to $\mu$ so that $\sum c_i=1$. 

For $\eta_i=1$ and $\sum c_i=1$, the conditional probability for $m$ input photon and $k$ detector output, $P(k|n)$, can be evaluated as
\begin{eqnarray}
P(k|m)&&=\sum_{l_1, l_2,\cdots, l_N=0}^{m} A_{l_1}^N c_1^{l_1} \times A_{l_2}^{N-l_1} c_2^{l_2}\times \cdots A_{l_N}^{l_N} c_N^{l_N}\\ \nonumber
&&=\sum_{l_1,l_2,\cdots,l_N=0}^{m} m!\times \frac{c_1^{l_1}}{l_1!}\times \frac{c_2^{l_2}}{l_2!} \times \cdots \times \frac{c_N^{l_N}}{l_N!}
\end{eqnarray}
where $\{l_1, l_2, \cdots, l_N\}$ has $k$ non-zero terms and $\sum l_i = n$ . This expression can be evaluated numerically with $O(m^N)$ complexity, which is tractable for a 4-element detector. In the experiment, $c_i$'s were characterized by measuring the counting distribution in the single-photon regime (see Fig.~\ref{fig:counting_ratio} for the measurement of $c_i$'s). 

\begin{acknowledgments}
We thank Jim Daley and Mark Mondol for technical support in nanofabrication; Changchen Chen and Franco Wong for allowing us to use their electro-optic modulator; and Emily Toomey and Brenden Butters for fruitful discussion. This research was supported by the Air Force Office of Scientific Research (AFOSR) grant under contract NO. FA9550-14-1-0052; National Science Foundation (NSF) grant under contract NO. ECCS-1509486; the Army Research Office (ARO) under Cooperative Agreement Number W911NF-16-2-0192. The views and conclusions contained in this document are those of the authors and should not be interpreted as representing the official policies, either expressed or implied, of the Army Research Office or the U.S. Government. The U.S. Government is authorized to reproduce and distribute reprints for Government purposes notwithstanding any copyright notation herein. D.Z. is supported by a National Science Scholarship from A*STAR, Singapore. H. C. is supported in part by a Samsung Scholarship. T.-J.L is supported by the Department of Defense National Defense Science and Engineering Graduate Fellowship. A.E.D. is supported by a National Aeronautics and Space Administration Space Technology Research Fellowship (award no. NNX14AL48H). 
\end{acknowledgments}

\section*{Contributions}
D.Z., Q.-Y.Z., K.K.B., and D.R.E. conceived the idea. D.Z. designed the device. D.Z. and T.-J.L. fabricated the device. D.Z., Q.-Y.Z., and H.C. performed the measurement. D.Z. and A.E.D. deposited the superconducting film. K.K.B. and D.R.E. supervised the project. All the authors discussed the results and wrote the manuscript.

\section*{Competing interests}
The authors declare no competing financial interests.

%

\widetext
\clearpage
\begin{center}
	\textbf{\large Supplementary Information
	}
\end{center}
\setcounter{equation}{0}
\setcounter{figure}{0}
\setcounter{table}{0}
\setcounter{page}{1}
\makeatletter
\renewcommand{\theequation}{S\arabic{equation}}
\renewcommand{\thefigure}{S\arabic{figure}}
\renewcommand{\bibnumfmt}[1]{[S#1]}
\renewcommand{\citenumfont}[1]{S#1}
\setcounter{section}{0}


\section{Device fabrication}

\begin{figure}[H]
	\centering
	\includegraphics[width = 0.9\textwidth]{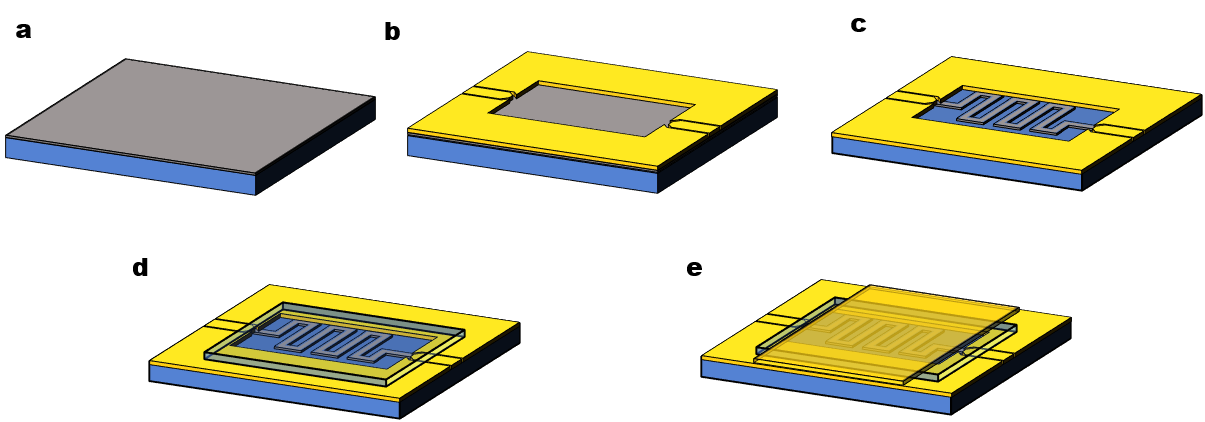}
	\caption[Fabrication process.]{\textbf{Fabrication process.} \textbf{a}, NbN deposition on a AlN-on-sapphire substrate. The NbN film was deposited using DC magnetron sputtering at 840 \degree C~\cite{Najafi2015b}. \textbf{b}, Gold pad fabrication. The bottom electrical contact pad and alignment mark were defined using contact photolithography. A bilayer resist with PMGI SF9 and S1813 were used to facilitate lift-off. The metal (5 nm Ti/50 nm Au/5 nm Ti) were deposited using electron-beam evaporation. \textbf{c}, Nanowire patterning. The superconducting nanowires were patterned using electron-beam lithography (EBL). 4\% HSQ was spin coated to the sample at 4 krpm. A 125 keV EBL system (Elionix F125) was used to expose the resist. The beam current was 1 nA, and the dose was 3840 $\upmu$C/cm$^2$. The HSQ was developed in 25\% TMAH for 2 min and rinsed with DI water. The HSQ pattern was transfered to the NbN film using reactive ion etching with CF$_4$ chemistry (He:CF$_4$ 7 sccm: 15 sccm). The etching was at 10 mTorr, 50 W for 1 min 45 s. \textbf{d}, Dielectric spacer fabrication. Dow Corning FOX-16 was spin coated at 3 krpm and baked at 250 \degree C for 2 min. The intended area was exposed using EBL at 20 nA with a dose of 800 $\upmu$C/cm$^2$, then developed in CD-26 for 70 s followed by rinsing in DI water. We measured the thickness of the spacer to be 450 nm using a surface profiler (Dektak). \textbf{e}, Top grounding plane fabrication. The top grounding plane was fabricated using a similar process as \textbf{b}. It was designed to extend outside the dielectric spacer to make a contact to the bottom common ground.}
	\label{fig:fab_flow}
\end{figure}

\begin{figure}[H]
	\centering
	\includegraphics[width = 0.9\textwidth]{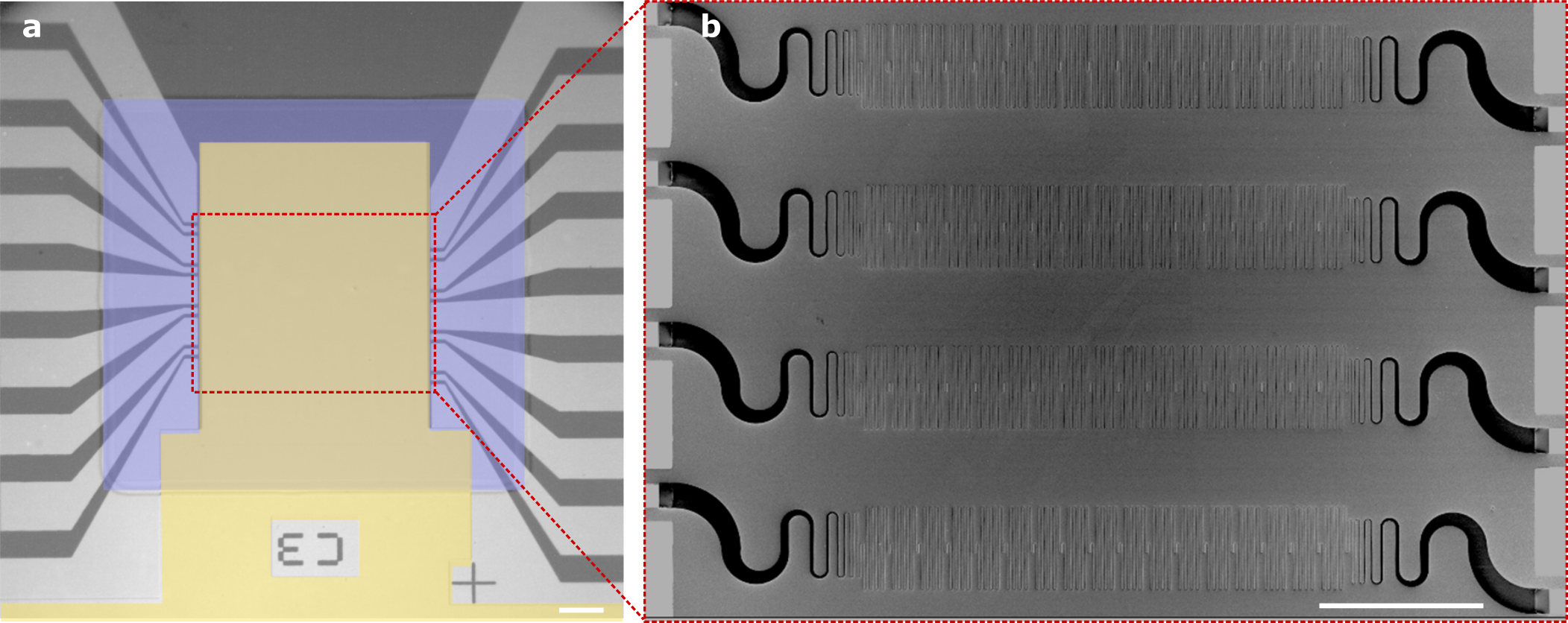}
	\caption[Micrographs of the fabricated device.]{\textbf{Micrographs of the fabricated device.} \textbf{a}, Optical micrograph of the final fabricated device. The light gray area on the bottom is the gold contact pads, and the dark gray area is the substrate. The purple area marks the middle dielectric spacer. The yellow area marks the top gold grounding plane. The red box at the center marks the actual location of the superconducting nanowire. \textbf{b}, Scanning electron micrograph of the superconducting nanowires before putting on the dielectric spacer and top ground. 4 sets of detectors were fabricated in the same device area. Scale bar: 100\,$\upmu$m.}
	\label{fig:device_micrograph}
\end{figure}

\section{Basic detector metrics of the 16-element device}
\subsection{Detector efficiency and dark count rate}
\begin{figure}[H]
	\centering
	\includegraphics[width = 0.8\textwidth]{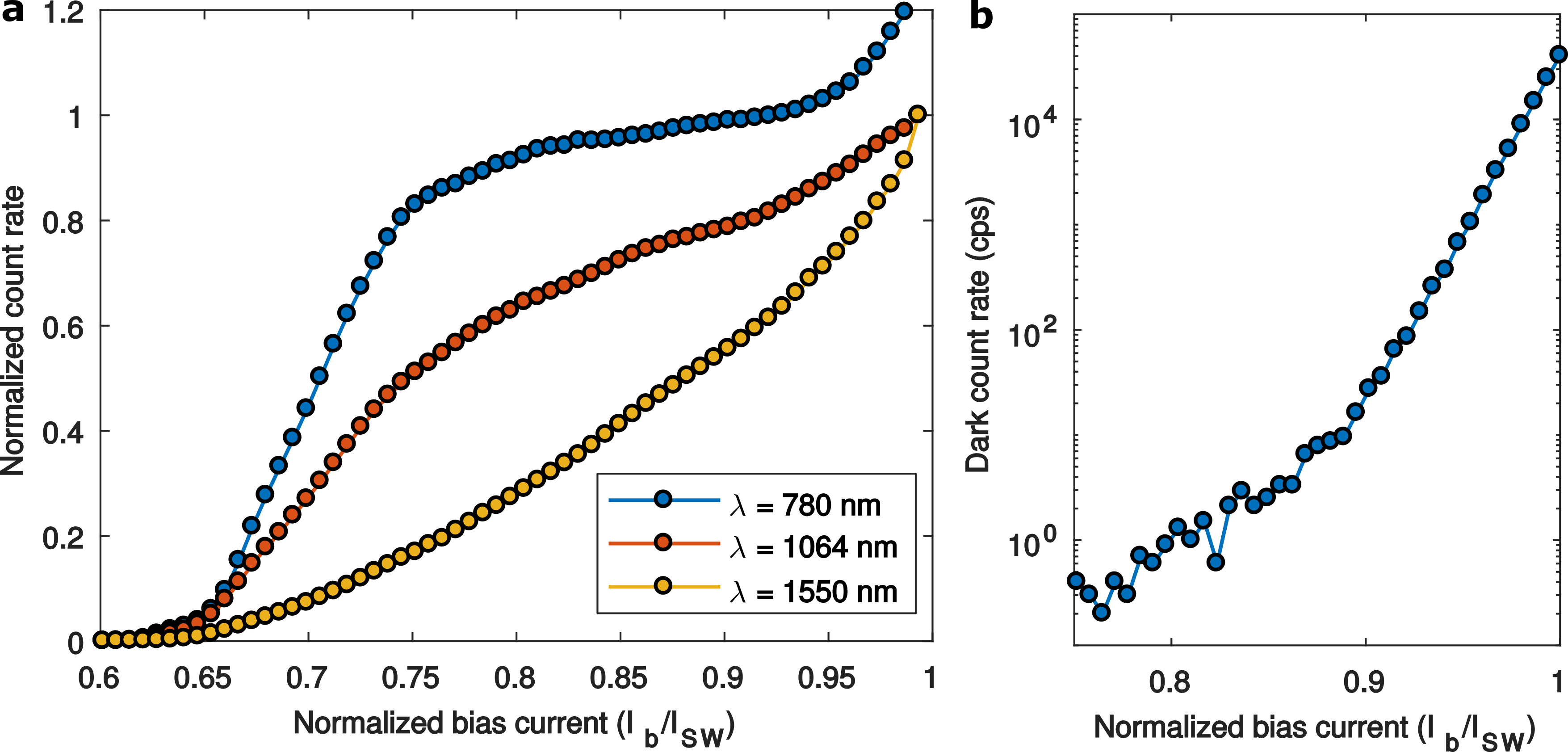}
	\caption[Detector efficiency and dark count in the 16-element detector]{\textbf{Detector efficiency and dark count in the 16-element detector} \textbf{a}, Normalized count rate vs. normalized bias current under different illumination photon energies. At 780\,nm illumination, the count rate was saturated, indicating high internal quantum efficiency. The increasing count rate near $I_\mathrm{SW}$ was likely due to the photon counts on the delay line. \textbf{b}, Dark count rate vs. normalized bias current.}
	\label{fig:DE_DCR}
\end{figure}

\subsection{Timing jitter} 
\begin{figure}[H]
	\centering
	\includegraphics[]{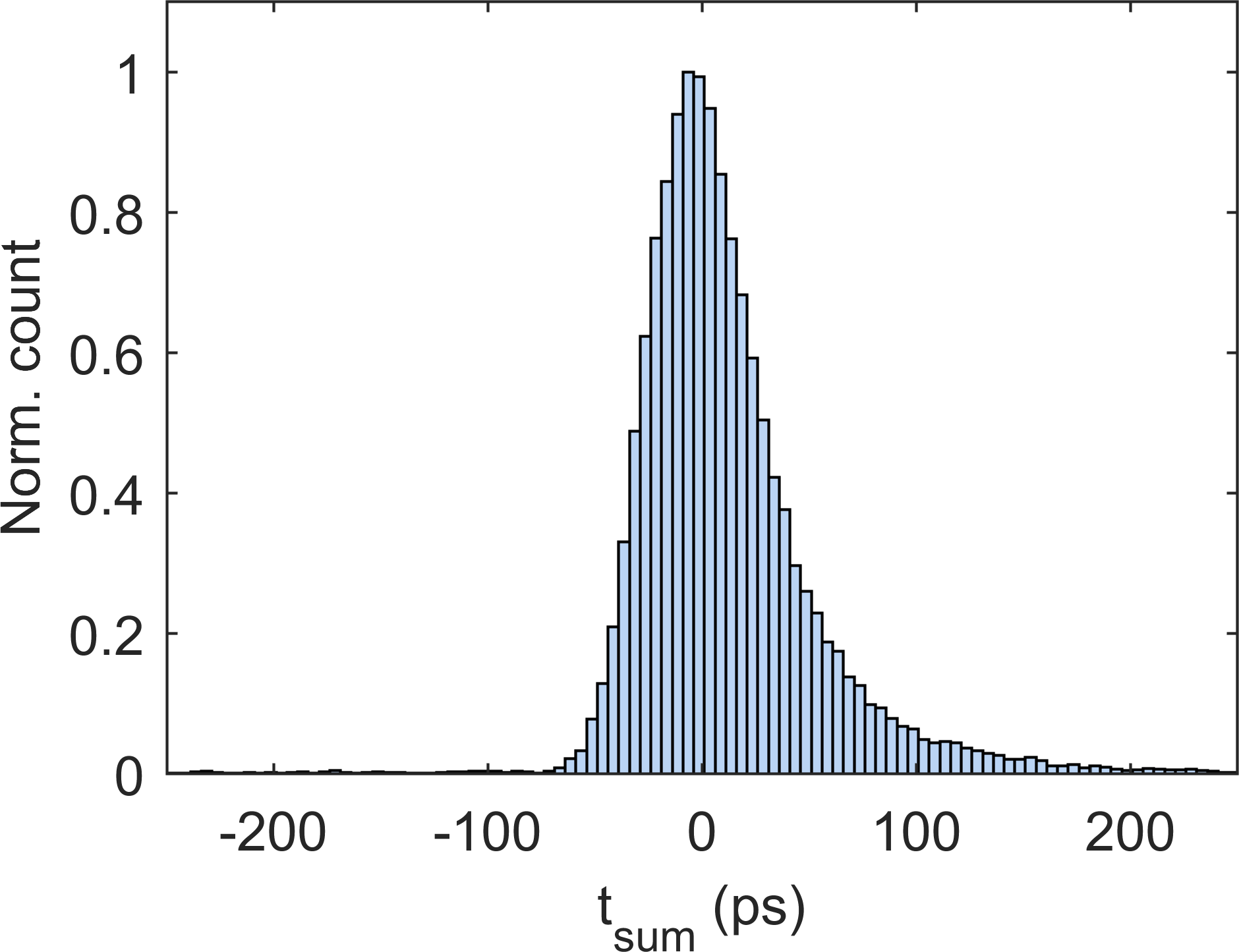}
	\caption[Sum jitter for the 16-element device.]{\textbf{Sum jitter for the 16-element device.} The $t_\mathrm{sum}$ was measured using the 1550\,nm sub-ps laser in the single-photon regime. The measured FWHM sum jitter was 59\,ps, which is higher than that in the 4-element detector (56\,ps). The increased jitter was likely due to the larger variation in pulse shapes (16 different pulse shapes), which causes fluctuation of triggering point when performing time tagging.}
	\label{fig:jitter_16_element}
\end{figure}

Fig.~\ref{fig:jitter_16_element} shows the distribution of $t_\mathrm{sum}$ of the 16-element detector measured using a 1550\,nm sub-ps pulsed laser in the single-photon regime, showing a FWHM sum jitter of 59\,ps. 

Device timing jitter, together with uncertainty of photon arrival time with respect to the reference clock, determines the minimum delay line required to resolve all the detection events. In the main text, we showed that the 4-element device had a FWHM difference jitter of 20\,ps and sum jitter of 56\,ps. For the 16-element device, the FWHM difference jitter was also 20\,ps and a sum jitter of 59\,ps (see Fig.~\ref{fig:jitter_16_element}.  This increased sum jitter was likely due to the variation in pulse shapes caused by imperfect impedance matching. The variation in pulse shape, especially on the rising edge, will induce fluctuation of triggering point for time tagging. Given these values, We estimated $\sigma_\mathrm{e}=20/2.355=8.5$\,ps and $\sigma_\mathrm{det}=(59^2-20^2)^{1/2}/2.355=23.6$\,ps. Based on the current timing jitter, for example, in order to resolve a pulsed spontaneous parametric down-conversion photon pair with a FWHM timing uncertainty of 2\,ps with respect to the timing reference given by the pump laser~\cite{Chen2017},  we would need a delay line of 150.6\,ps (6$\sigma$) to achieve $>$99.7\% fidelity.

\subsection{Maximum count rate}
\begin{figure}[H]
	\centering
	\includegraphics[]{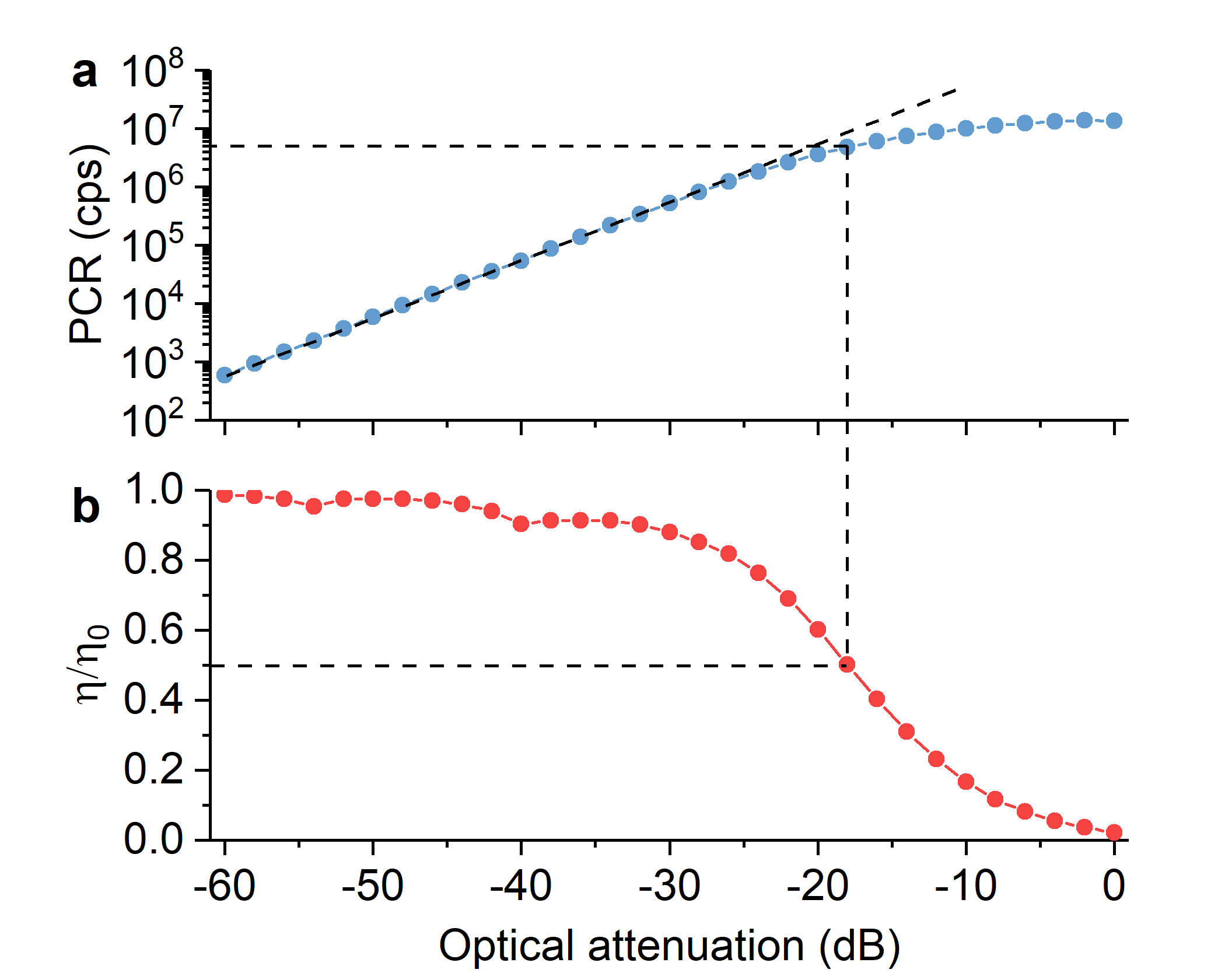}
	\caption[Maximum count rate of the 16-element detector]{\textbf{Maximum count rate of the 16-element detector.} \textbf{a}, Measured photon count rate (subtracted dark count rate) as a function of optical attenuation. \textbf{b}, Normalized detection efficiency as a function of optical attenuation. The maximum count rate was measured to be 4.8 MHz at the 3 dB suppression point for the detection efficiency. The light source was a superluminescent diode with a center wavelength at 1550 nm. We used the setup described in ~\textcite{Zhao2014a} to avoid capacitive charging at the amplifier.}
	\label{fig:maximum_CR}
\end{figure}

Figure~\ref{fig:maximum_CR} shows the count rate measurement of the 16-element detector. At 3 dB efficiency suppression point, the maximum count rate was 4.8 MHz.

The reset time of the detector is limited by the kinetic inductance of the nanowire~\cite{Kerman2006a}. The maximum count rate can be roughly estimated as $R_\mathrm{load}/[3N(L_\mathrm{det}+L_\mathrm{delay})]=O(1/N)$, where $N$ is the number of segments, $L_\mathrm{load}$ is the load impedance (50\,$\Omega$ in our case), $L_\mathrm{det}$ and $L_\mathrm{delay}$ are the inductance for each detector segment and each delay line.  One could circumvent the kinetic inductance limit by using an AC/pulsed bias. This method would require broadband, perfect impedance matching between the detector and readout circuits.

\section{Supplementary figures for the 4-element device}
\subsection{Counting ratio and laser pulse width estimation}
\begin{figure}[H]
	\centering
	\includegraphics[width = 0.8\textwidth]{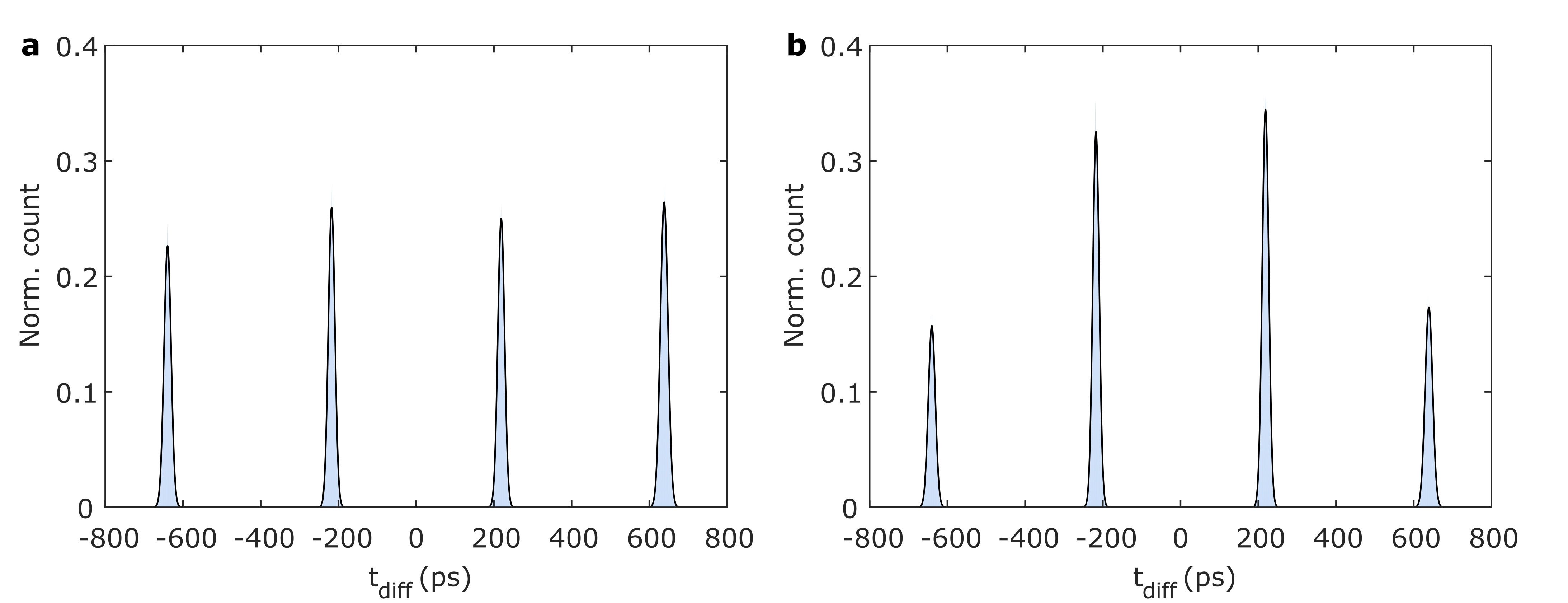}
	\caption[Counting rate distribution in the 4-element detector]{\textbf{Counting rate distribution in the 4-element detector.} \textbf{a}, When the probing fiber was far away from the detector, all the segments were illuminated uniformly, and they have a relatively uniform counting rate with a ratio of [0.2263, 0.2595, 0.2500, 0.2642]. \textbf{b}, When the fiber was focused at the center of the detector, the middle two segments had a higher counting rate, and the counting ratio was [0.1573, 0.3252, 0.3443, 0.1732]. This ratio was used as $c_i$ in the photon statistics modeling.}
	\label{fig:counting_ratio}
\end{figure}

\begin{figure}[H]
	\centering
	\includegraphics[]{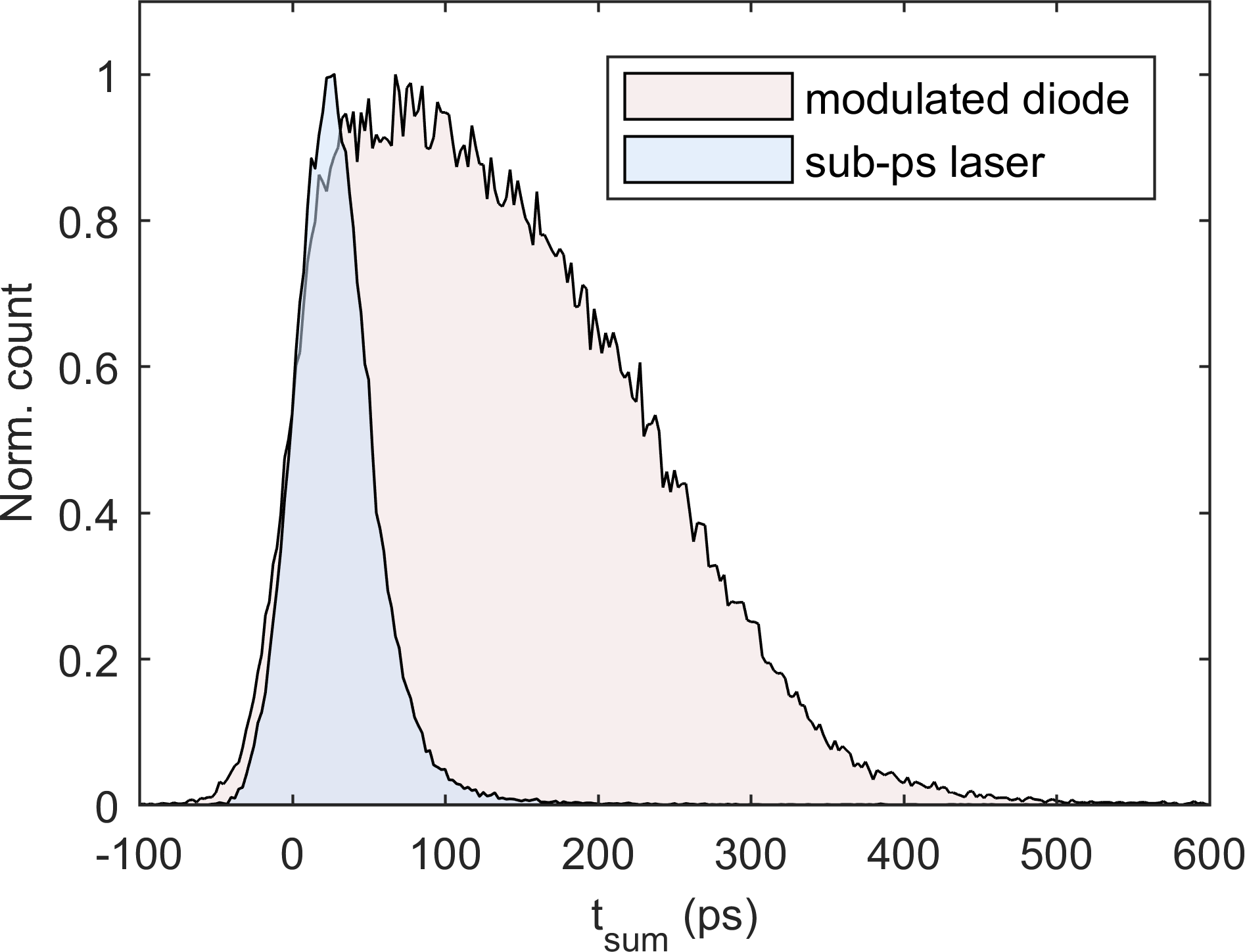}
	\caption[Laser pulse shape comparison.]{\textbf{Laser pulse shape comparison.} The FWHM sum jitter was 50\,ps when illuminated using the sub-ps laser, while it was 240\,ps when illuminated using the modulated diode laser.}
	\label{fig:laser_pulse_width}
\end{figure}

\subsection{Complete list of all pulse shapes and their fingerprints}

\begin{figure}[H]
	\centering
	\includegraphics[]{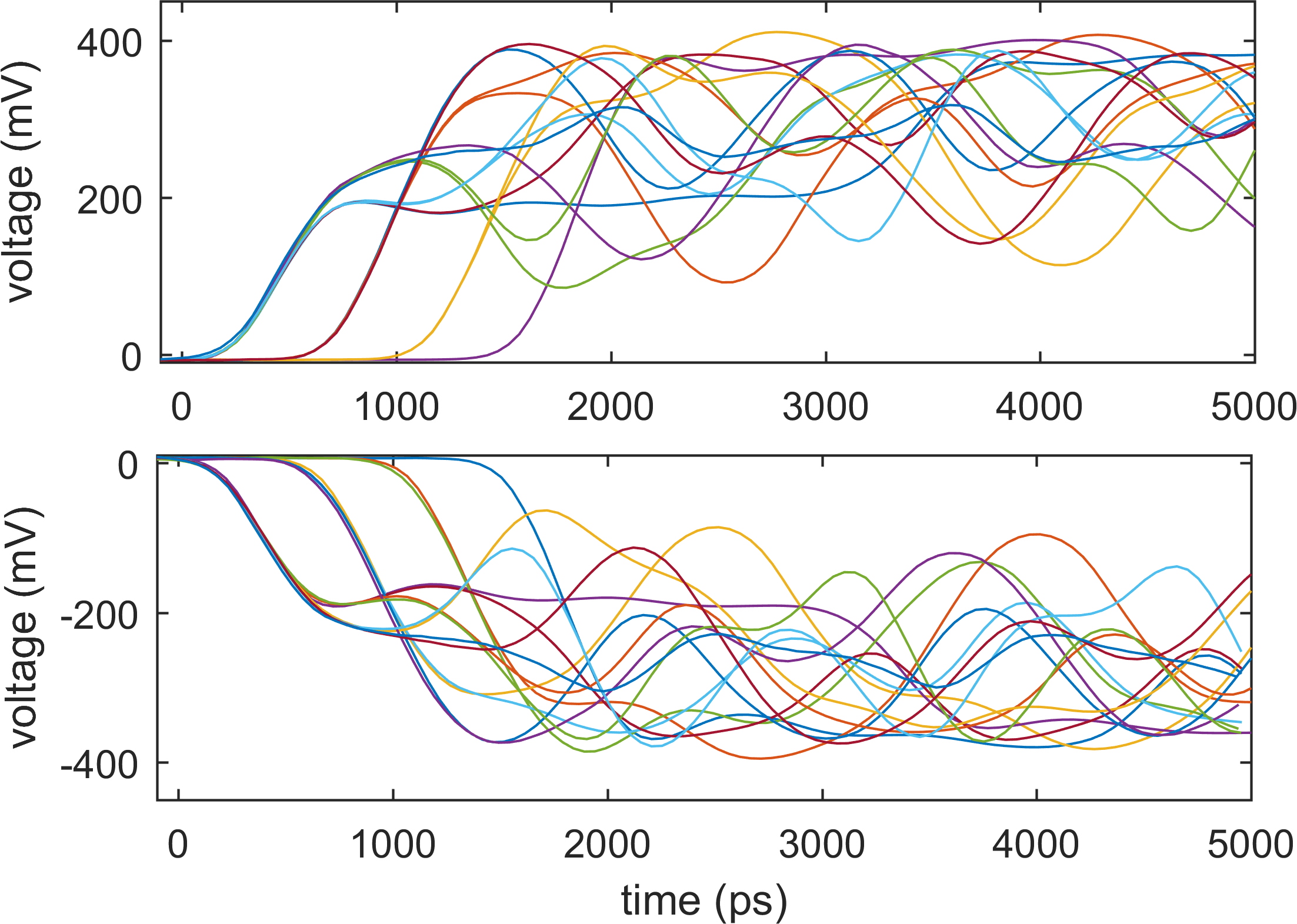}
	\caption[Average pulse shapes for all detection cases]{\textbf{Average pulse shapes for all detection cases.}}
	\label{fig:average_pulse_shape}
\end{figure}

\begin{figure}[H]
	\centering
	\includegraphics[]{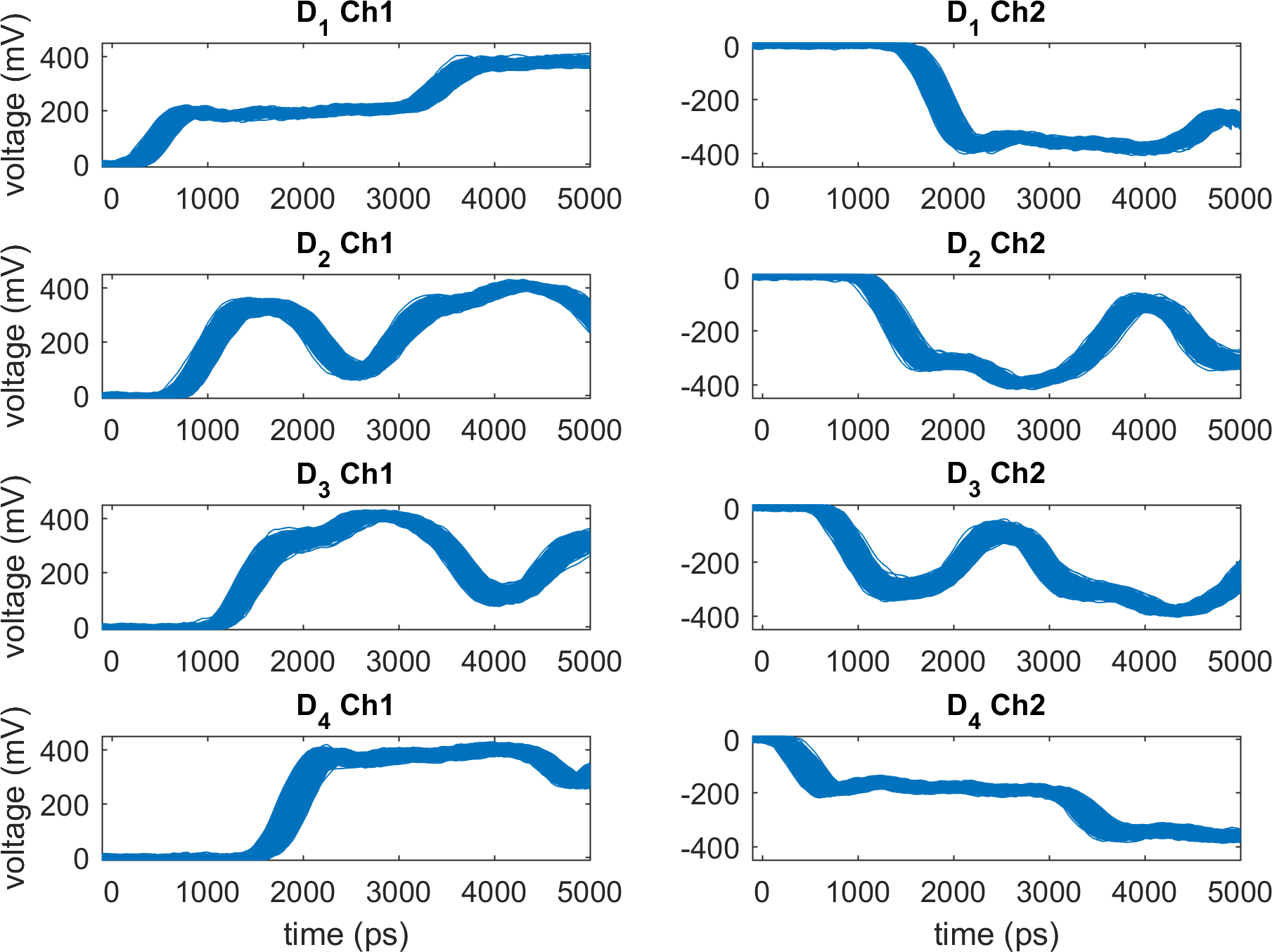}
	\caption[Detector pulses for all single-photon cases]{\textbf{Detector pulses for all single-photon cases.}}
	\label{fig:single_events}
\end{figure}

\begin{figure}[H]
	\centering
	\includegraphics[]{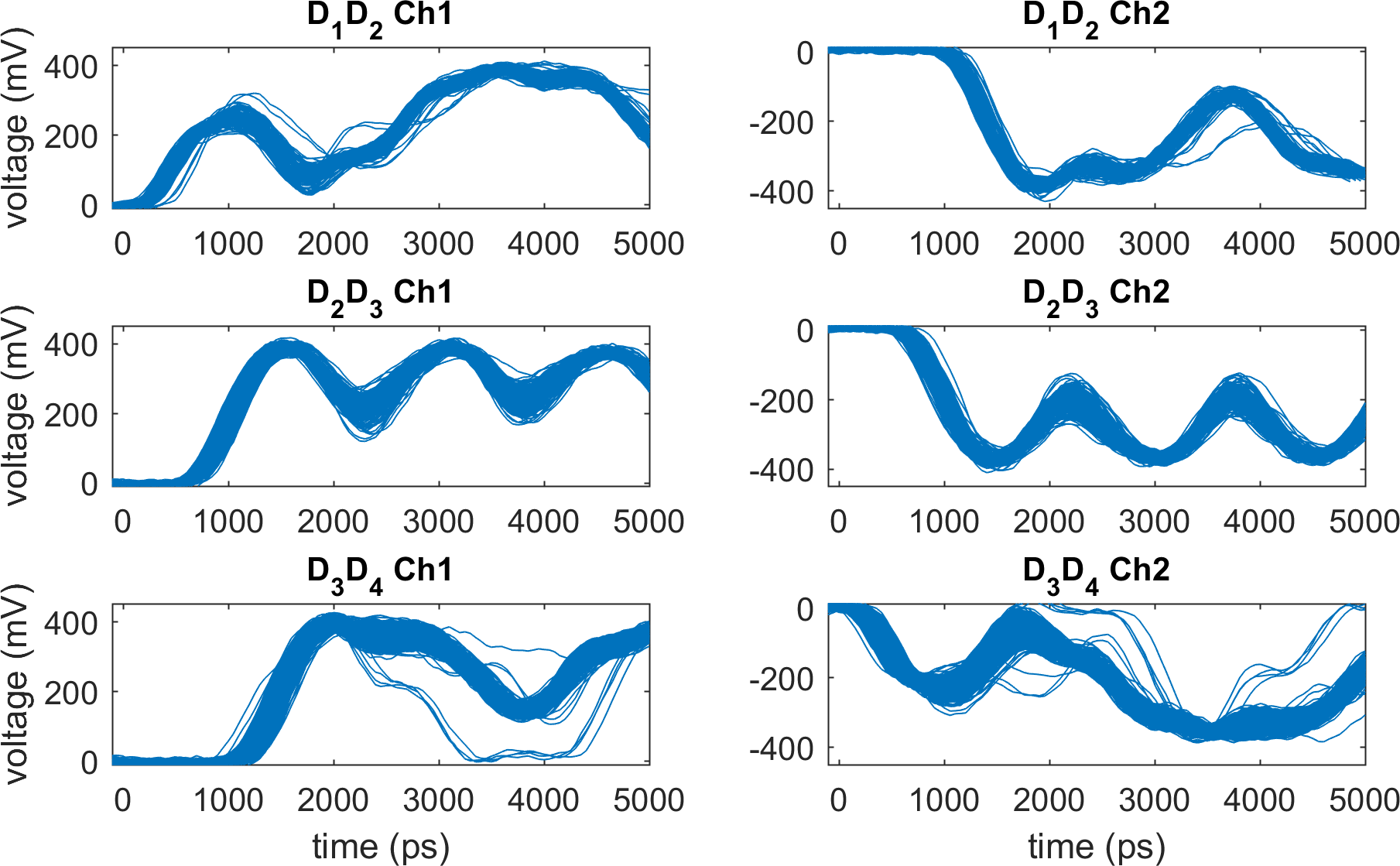}
	\caption[Detector pulses for unambiguous two-photon events]{\textbf{Detector pulses for unambiguous two-photon events.} These two-photon detection events originate from adjacent detectors, which has no unambiguity of hiding a three-photon events. We noticed some irregular pulse shapes for $D_1D_2$ and $D_3D_4$ events, but the reason was not completely understood yet. }
	\label{fig:umambiguous_two_photon_events}
\end{figure}

\begin{figure}[H]
	\centering
	\includegraphics[width = .8\textwidth]{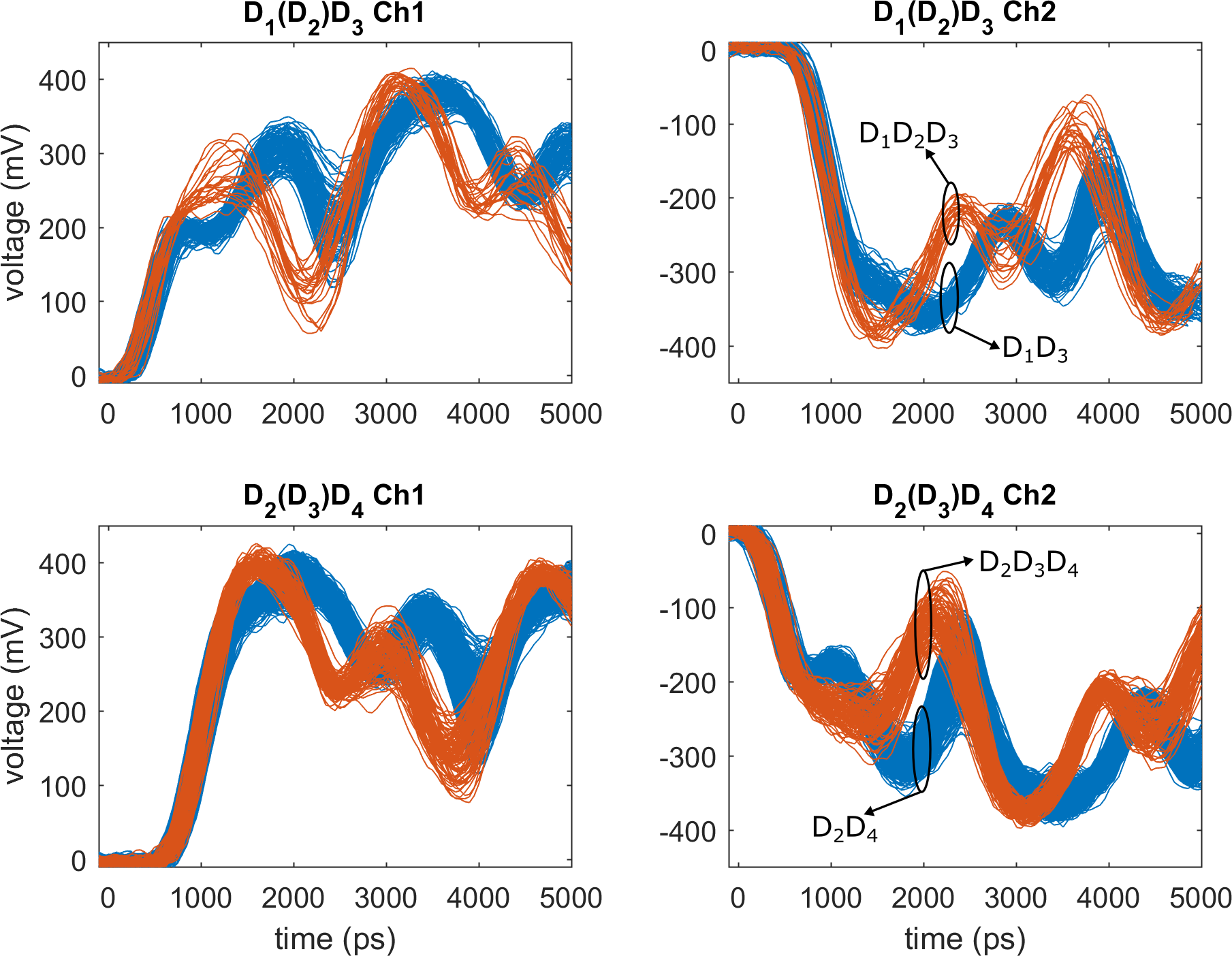}
	\caption[Detector pulses for ambiguous two-photon events]{\textbf{Detector pulses for ambiguous two-photon events.} The circles indicate the fingerprints used to distinguish the events.}
	\label{fig:ambiguous_two_photon_events}
\end{figure}

\begin{figure}[H]
	\centering
	\includegraphics[width = .8\textwidth]{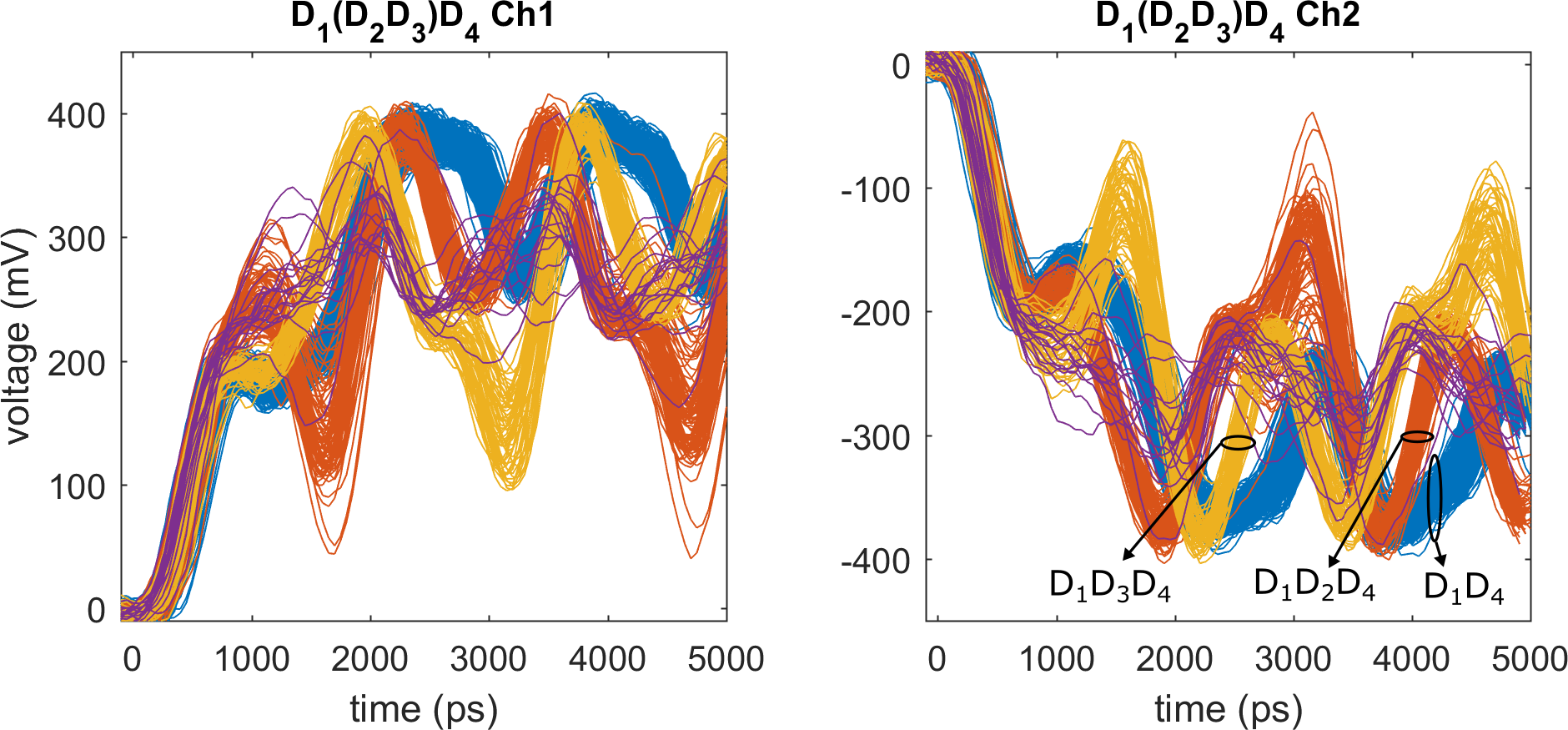}
	\caption[Detector pulses for ambiguous two, three, and four events]{\textbf{Detector pulses for ambiguous two, three, and four events.} The circles indicate the fingerprints used to distinguish the events. We counted $D_1D_2D_3D_4$ events (purple) by identifying pulses that do not match the other three fingerprints. }
	\label{fig:three_four_events}
\end{figure}

\section{Discussion on waveguide integration }

\begin{figure}[H]
	\centering
	\includegraphics{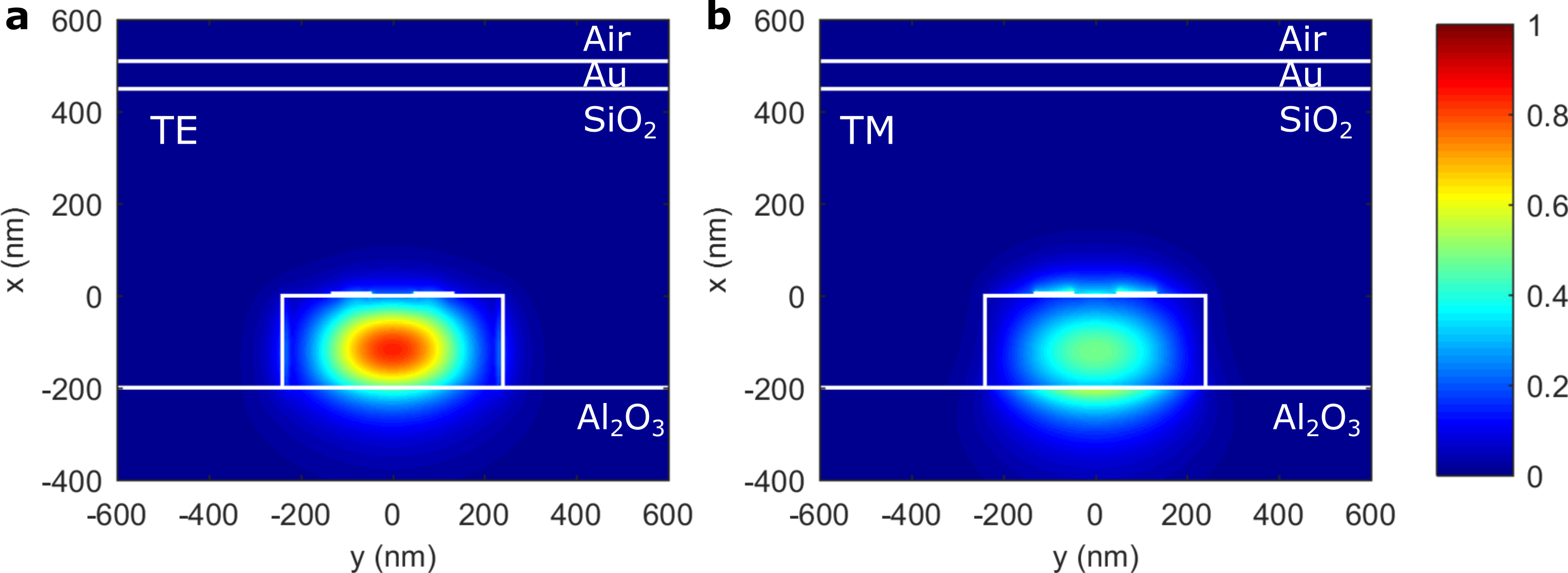}
	\caption[Calculation of the optical absorption in the superconducting nanowire when integrated on waveguide]{\textbf{Calculation of the optical absorption in the superconducting nanowire when integrated on waveguide.} The mode pattern  (normalized $|\mathbf{E}|^2$) for a 450 nm $\times$ 200 nm AlN waveguide with integrated 2-SNAPs at 637 nm wavelength. \textbf{a}, Transverse electric mode. $n_\mathrm{eff}=1.86+0.0134i$, and absorption rate is 1.146 dB/$\upmu$m. \textbf{b}, Transverse magnetic mode. $n_\mathrm{eff}=1.83+0.0085i$, and absorption rate is 0.73 dB/$\upmu$m. The simulation was performed using Lumerical MODE Solutions.}
	\label{fig:waveguide_integration}
\end{figure}

The detection efficiency is the product of the internal quantum efficiency ($\eta_\mathrm{int}$) and optical absorption ($\eta_\mathrm{abs}$). The optical absorption can in principle reach unity when the detector is integrated on an optical waveguide~\cite{Hu2009a,Pernice2012}. Simply etching the AlN substrate into a 450\,nm$\times$200\,nm ridge waveguide, the 80-nm-wide 2-SNAP will have an absorption rate of 1.15\,dB/$\upmu$m for the transverse electric (TE) mode at 637\,nm wavelength, which corresponds to the zero-phonon line of nitrogen vacancy centers in diamond. Figure~\ref{fig:waveguide_integration} shows the numerical simulation for the waveguide mode and absorption rate. To achieve $>$90\% absorption, the 2-SNAP needs to be 8.7\,$\upmu$m long. Adding a reflector or photonic crystal cavity on the waveguide can further reduce the length~\cite{Akhlaghi2015}. The ability to control absorption by changing the nanowire length can be used to cascade multiple partially absorbing detector chains in parallel on an optical waveguide array. This method can be used to handle input states with more than two photons or resolving multi-photons in the same mode/waveguide. 

\end{document}